\documentstyle[11pt]{article}
\newcommand{\be}{\begin{equation}}
\newcommand{\ee}{\end{equation}}
\newcommand{\bea}{\begin{eqnarray}}
\newcommand{\eea}{\end{eqnarray}}
\newcommand{\nn}{\nonumber}
\newcommand{\p}{\phi}
\newcommand{\pp}{\tilde \phi}
\newcommand {\pb}{\bar \phi}

\begin{document}
 \title{Non-linear Laplace equation, de Sitter vacua and
information geometry}
 \author{Farhang Loran\thanks{e-mail:
loran@cc.iut.ac.ir}\\ \\
  {\it Department of  Physics, Isfahan University of Technology (IUT)}\\
{\it Isfahan,  Iran}}
\date{}
  \maketitle
 \begin{abstract}
 Three exact solutions say $\phi_0$ of massless scalar theories on Euclidean space, i.e.
 $D=6\ \phi^3$, $D=4\ \phi^4$ and $D=3\ \phi^6$
 models are obtained which share similar properties. The information geometry of their
 moduli spaces coincide with the Euclidean $\mbox{AdS}_7$, $\mbox{AdS}_5$ and
 $\mbox{AdS}_4$ respectively on which $\p_0$ can be described as a stable
 tachyon. In $D=4$ we recognize that the SU(2) instanton density is proportional to
 $\phi_0^4$. The original action $S[\phi]$ written in terms of new scalars
 ${\tilde \phi}=\phi-\phi_0$ is shown to be equivalent to an interacting
 scalar theory on $D$-dimensional de Sitter background.
 \end{abstract}
 AdS/CFT correspondence \cite{Mal}, as a bulk/boundary correspondence, is a quantitative
 realization of the holographic principle. In \cite{Witten} Witten showed that
 the metric on the boundary of the AdS space is well-defined only up to a
 conformal transformation and the correlation functions of the CFT on the boundary are
 given by the dependence of the supergravity action on the asymptotic behavior at
 infinity,  see also \cite{Reh}.
 Using the metric $ds^2=\frac{1}{{x^0}^2}(d{x^0}^2+d
 {x^1}^2+\cdots+d {x^d}^2)$ for the Euclidean $\mbox{AdS}_{d+1}$ Witten showed that the
 generating function for CFT correlators, $I[\phi]=\ln \left<\exp \int \p
 {\cal O}\right>$ is
 \be
 I[\phi]=\int d^dy d^dz
 \frac{\phi_0({\vec y})\phi_0({\vec z})}{\left|{\vec y}-{\vec
 z}\right|^{2(d+\lambda_+)}},
 \label{generator}
 \ee
 where $\p_0$ here, is some scalar field on the boundary, determined by the asymptotic
 behavior of scalar fields $\Phi$ in the bulk: $\Phi\sim {x^0}^{-\lambda_+}\p_0$
 as $x^0\to 0$. Here, $\lambda_+$ is the larger root of the equation
 $\lambda(\lambda+d)=m^2$. These results led us to a classical
 interpretation for EAdS/CFT correspondence as the
 relation between the solutions of the Klein-Gordon equation $\p(x^0,{\vec x})$ (bulk fields)
 and the Cauchy data $\p(0,{\vec x})$ (boundary fields) \cite{map}.
 In fact under the conformal
 transformation $\delta_{\mu\nu}\to g_{\mu\nu}={x^0}^{-2}\delta_{\mu\nu}$ that gives the
 $\mbox{EAdS}_{d+1}$ metric mentioned above in terms of $\delta_{\mu\nu}$, the metric of the
 $D$-dimensional flat Euclidean space $R^{d+1}$, massless fields $\p$
 on $R^{d+1}$ transform to massive scalars $\Phi={x^0}^{-\lambda_+}\p$ with mass
 $-\frac{d^2-1}{4}$. From the classical equation of motion $\delta S[\p]=0$ one can determine
 $\p(x^0,{\vec x})$ in terms of the Cauchy data $\p_0({\vec x})=\p(0,{\vec x})$.
 Inserting the solution in $S[\p]$ one obtains $I[\p]$ given in Eq.(\ref{generator}).
 By the same method, though only for scalars with specific mass
 $m^2=\frac{d^2-1}{4}$,  the correlation functions of the boundary operators
 in $\mbox{dS}_{d+1}/\mbox{CFT}_d$ correspondence that Strominger  \cite{Strom} explicitly
 calculated for $d=2$ and proposed for general $d$ can be
 obtained \cite{map}. By generalizing the method to free spinors the boundary
 term to be added to the bulk Dirac action necessary for AdS/CFT
 and dS/CFT correspondence \cite{Hen} are obtained for general free massive spinors in
 (A)dS space \cite{Spin}.
 \par
 What can one learn about AdS/CFT  correspondence if one uses this
 method  for interacting scalar theories instead of free scalars?
 The scalar field theories that can be considered are massless $D=6\ \phi^3$, $D=4\ \phi^4$
 and $D=3\ \phi^6$ models \cite{Spin,phi4} given by the action,
 \be
 S[\p]=\int d^Dx
 \left(\frac{1}{2}\delta^{\mu\nu}\partial_\mu\p\partial_\nu\p-
 \frac{g}{\left(\frac{2D}{D-2}\right)}\p^{\frac{2D}{D-2}}\right),\hspace{1cm}D=3,4,6,
 \label{action}
 \ee
 which is classically invariant under rescaling $
 x\to \lambda x,\ \p\to \lambda^\frac{2-D}{2}\p.$
 Using the conformal transformation  from  $R^D$
 (D-dimensional Minkowski space-time) to $\mbox{EAdS}_D$
 ($\mbox{dS}_D$) one obtains the same interacting theory, i.e. $\Phi^\frac{2D}{D-2}$-model
 but for  scalars with mass $m^2=\pm\frac{d^2-1}{4}$ on $D=d+1$-dimensional (A)dS
 space. The corresponding generating function for boundary CFT
 correlators can be obtained by solving (by perturbation) $\p(x^0,{\vec x})$
 in terms the Cauchy data and inserting the solution in $S[\p]$
 \cite{phi4}. As is discussed in \cite{phi4} finding the solution of the equation of motion
 as power series in $g$ the coupling constant,
 is necessary for applicability of the above mentioned method.
 As we will see, there are exact solutions of the corresponding non-linear Laplace  equation
 which although are not useful for that purpose but open a new window to the AdS/CFT
 correspondence for critical scalar field theories
 (\ref{action}).
 \section*{The Information Geometry}
 As is shown in  \cite{phi4}, critical scalar field theories (\ref{action}) are particular
 in the sense that the $SO(D)$-invariant (in flat Euclidean space) nonlinear
 Laplace equation,  $\nabla^2 \p+g\p^n=0,$ $(g>0)$, in which $\nabla^2=
 \delta^{\mu\nu}\partial_\mu\partial_\nu$, has solutions like
 \be
 \p_0(s)=\frac{\alpha}{\left(\beta^2+(s-a)^2\right)^\gamma},\hspace{1cm}
 (s-a)^2=\delta_{\mu\nu} (x-a)^\mu(x-a)^\nu.
 \label{soliton}
 \ee
 for some constants $a^\mu$, $\alpha$, $\beta$ and $\gamma$ {\bf only if}
 $n=\frac{2D}{D-2}-1$ and
 $D=3,4,6$. In these cases $\gamma=\frac{D-2}{2}$, and
 \be
 \beta^2=\frac{g}{D(D-2)}\alpha^{\frac{4}{D-2}},
 \ee
 which for example for $D=4$ gives,
 \be
 \p_0\sim\frac{\beta}{\beta^2+(x-a)^2}.
 \ee
 We note that $\p_0(x;\beta,a_\mu)$ is
 invariant under rescaling, see appendix A and
 is a stable classical solution of an unstable model ($V(\p)\sim-\phi^4$).
 An interesting observation is that in $D=4$, the SU(2) instanton
 density is \cite{Blau}
 \be
 \mbox{tr} F^2=96 \frac{\beta^4}{(\beta^2+(x-a)^2)^4}\sim\p_0^4.
 \label{density}
 \ee
 In \cite{Blau} $\beta$ in Eq.(\ref{density}) is considered as the size of the
 instanton, suggesting to call $\beta$ in Eq.(\ref{soliton}) the size of $\p_0$.
 Considering $\theta^I=\beta,a^\mu$, $I=0,\cdots,D$ in Eq.(\ref{soliton}) as moduli,
 the Hitchin information metric of the moduli space, defined as follows \cite{Hit}:
 \be
 {\cal G}_{IJ}=\frac{1}{N(D)}\int d^Dx
 {\cal{L}}_0\partial_I\left(\log{\cal{L}}_0\right)
 \partial_J\left(\log{\cal{L}}_0\right),
 \ee
 can be shown to describe Euclidean $\mbox{AdS}_{D+1}$ space:
 \be
 {\cal G}_{IJ}d\theta^Id\theta^J=\frac{1}{\beta^2}\left(d\beta^2+d
 a^2\right).
\label{Mod-met}
 \ee
 $N(D)$ is a normalization constant,
 \be
 N(D)=\frac{D^3}{D+1}\int d^D x {\cal L}_0,
 \label{N}
 \ee
 and
 \be
 {\cal{L}}_0=-\frac{1}{2}\p_0\nabla^2\p_0-\frac{g}{\left(\frac{2D}{D-2}\right)}
 \p_0^{\frac{2D}{D-2}}
 =\frac{g}{D}\p_0^{\frac{2D}{D-2}},
 \label{L0}
 \ee
 is the Lagrangian density calculated at $\p=\p_0$. See appendix B
 for details. Similar results are obtained for the information geometry of instantons
 on $R^4$, for ${\cal N}=\frac{1}{2}$ U(N) theories and for instantons on noncommutative
 space. See for example \cite{Blau,Rey}.
 \par
 $\p_0$ as a function of $\theta^I$'s is a free stable-tachyon field
 on ${\mbox{EAdS}_{D+1}}$
 as it satisfies the Klein-Gordon equation given in terms of the metric (\ref{Mod-met}),
 \be
 \left(\beta^2\partial_\beta^2+(1-D)\beta\partial_\beta+\beta^2\partial_a^2+
 \frac{D^2-4}{4}\right)
 \p_0=0.
 \ee
 The tachyon is stable as far as $\frac{-D^2}{4}<m^2<0$
 \cite{Witten}.
 \par
 $\p_0$ as a function of $g$ the coupling constant (or $\beta^2$), can not be
 analytically continued to $g=0$. For $g=0$, $\p_0$ is the
 Green function of the Laplacian  operator i.e. $\nabla^2\p(x,a)=\delta^D(x-a)$
 and does not satisfy the Klein~Gordon equation $\nabla^2\p=0$ for free scalar theory.
 This shows that $\p_0$ can not be
 obtained by perturbation around $g=0$. In \cite{Blau}, the
 same asymptotic behavior for the instanton density is observed
 and  $\mbox{tr}F^2$ is interpreted as the {\em boundary to bulk}
 propagator of a massless scalar field on ${\mbox AdS}_5$.
 \section*{$\p_0$ as classical de Sitter vacua}
 Rewrite the action (\ref{action})
  in terms of new fields $\pp=\p-\p_0$, one obtains
 \be
 S[\p]=S[\p_0]+S_{\mbox{free}}[\pp]+S_{\mbox{int}}[\pp],
 \label{act}
 \ee
 where $S[\p_0]=\int d^Dx {\cal L}_0$ (see Eq.(\ref{L0})),
 \be
 S[\p_0]=
 \frac{D^{\frac{D-2}{2}}(D-2)^{\frac{D}{2}}\pi^{\frac{D+1}{2}}}{2^{D-1}\Gamma(\frac{D+1}{2})}
 g^{\frac{2-D}{2}}=
 \left\{\begin{array}{ccc}\frac{192\pi^3}{5g^2}&&D=6,\\\\\frac{8\pi^2}{3g}&&D=4,\\\\
 \frac{\pi^2}{4}\sqrt{\frac{3}{g}}&&D=3\end{array}\right.
 \ee
 and
 \be
 S_{\mbox{free}}[\pp]=\int d^D x \left(
 \frac{1}{2}\delta^{\mu\nu}\partial_\mu\pp\partial_\nu\pp+\frac{1}{2}M^2(x)\pp^2\right)
 \label{free}
 \ee
 in which,
 \be
 M^2(x)=-g \left(\frac{D+2}{D-2}\right)\p_0^{\frac{4}{D-2}}
 =-(2+D)D\frac{\beta^2}{\left(\beta^2+(x-a)^2\right)^2}.
 \label{mass}
 \ee
 Defining $\pb=\Omega^{\frac{2-D}{4}}\pp$,
 one can show that $S_{\mbox{free}}[\pp]$ given in Eq.(\ref{free})
 is the action of the scalar
 field $\pb$ on some conformally flat background
 with metric $g_{\mu\nu}=\Omega\delta_{\mu\nu}$:
 \be
 S_{\mbox{free}}[\pp]=\int d^D x \sqrt{\left|g_{\mu\nu}\right|}
 \left(\frac{1}{2}g^{\mu\nu}\partial_\mu\pb
 \partial_\nu\pb+\frac{1}{2}(\xi R+m^2)\pb^2\right).
 \label{Curvedaction}
 \ee
 Here, $m^2\Omega=M^2(x)$, where $m^2$ is the mass of $\pb$
 (undetermined) and $M^2(x)$ is given in Eq.(\ref{mass}). $R$ is the curvature scalar
 and $\xi=\frac{D-2}{4(D-1)}$ is the conformal coupling constant.
 This result is surprising as one can show that the Ricci tensor
 $R_{\mu\nu}=\Lambda_D g_{\mu\nu}$, where
  \be
  \Lambda_D=-m^2\frac{4(D-1)}{D(D+2)}.
 \label{Lambda}
  \ee
  Since $-m^2>0$ as far as $\Omega>0$, one verifies that
  $\Lambda_D>0$ which means that $\pb$ lives on $D$-dimensional
  de~Sitter space which radius is proportional to $-m^{-2}$. See
  Appendix C for details.
  \par
  The interacting part of the action,
  $S_{\mbox{int}}[\pp]=\int d^Dx\sqrt{\left|g_{\mu\nu}\right|}{\cal L}_{\mbox{int}}$
  is well-defined in terms of $\pb$ on the corresponding $\mbox{dS}_D$:
  \be
  {\cal L}_{\mbox{int}}=\left\{\begin{array}{lll}
   \frac{-g}{3}\pb^3,&&D=6,\\\\
  -g\sqrt{\frac{-m^2}{3g}}\pb^3-\frac{g}{4}\pb^4,&&D=4,\\\\
  -\frac{10}{3}g\left(\frac{-m^2}{5g}\right)^{\frac{3}{4}}\pb^3
  -\frac{5}{2}g\left(\frac{-m^2}{5g}\right)^{\frac{1}{2}}\pb^4
 -g\left(\frac{-m^2}{5g}\right)^{\frac{1}{4}}\pb^5
  -\frac{g}{6}\pb^6,&&D=4.\end{array}\right.
  \ee
  It is interesting to note that in $D=4$, by a shift
  of the scalar field $\pb\to\pb-\sqrt{\frac{-m^2}{3g}}$ the action
  (\ref{act}) can be written in the $\mbox{dS}_4$ as follows:
  \be
  S[\pb]=\int d^D x \sqrt{\left|g_{\mu\nu}\right|}
 \left(\frac{1}{2}g^{\mu\nu}\partial_\mu\pb
 \partial_\nu\pb+\frac{1}{2}(\xi
 R)\pb^2-\frac{g}{4}\pb^4\right)+
 \int d^D x \sqrt{\left|g_{\mu\nu}\right|}\left(\frac{-m^4}{36
 g}\right).
  \ee
 \section*{$\p_0$ on Minkowski space-time}
 After a Wick rotation $x^0\to ix^0$, $\p_0$ given in
 (\ref{soliton}) can be shown to satisfy the corresponding non-linear wave equation
 on Minkowski space-time. These solutions have a time-like
 singularity. The singularity is a hypersurface given by the
 equation $-(x^0-a^0)^2+(x^1-a^1)+\cdots+(x^D-a^D)^2=-\beta^2$,
 which can be considered as a $D-1$ dimensional anti de Sitter
 space. \footnote{The singularity becomes space-like i.e. a $\mbox{dS}_{D-1}$ hypersurface
 if the coupling $g$ is negative.} The idea, here, is to some extend similar
 to the holographic reduction of Minkowski space-time \cite{Sol} where the Minkowski
 space-time is sliced in terms of Euclidian AdS and Lorentzian dS slices which correspond
 to the time-like and space-like regions respectively.
 Considering
 only the free part of the scalar action
 $S_{\mbox{free}}(\pp)$ given in Eq.(\ref{free}), one can verify
 that $D$-dimensional free scalar theory given by
 $S_{\mbox{free}}(\pp)$, induces a free (but unstable) scalar
 theory on the AdS hypersurface, the singularity. To show this,
 first note that the equation of motion for the scalars $\pp$
 is
 \be
 \left(\Box+\frac{D(D+2)\beta^2}{(\beta^2+x^2)^2}\right)\pp=0,\hspace{1cm}x^2=-{x^0}^2+{\vec
 x}^2,
 \label{a1}
 \ee
 where without losing the generality we have assumed $a^\mu=0$.
 Defining new coordinates
 \bea
 x^0&=&(R+\beta)\cosh{\rho},\nn\\
 x^i&=&(R+\beta)\sinh{\rho}\
 z_i,\hspace{1cm}\sum_{i=1}^{D-1}z_i^2=1,
 \eea
 which locates the singularity at $R=0$, Eq.(\ref{a1}) can be written as follows,
 \be
 \left(\Box_R+\frac{1}{R^2}\Box_{\rho,z_i}+\frac{D(D+2)\beta^2}{R^2(R+2\beta)^2}\right)
 \pp(R;\rho,z_i)=0.
 \label{a2}
 \ee
 The ansatz for scalar fields living on the singularity is
 $\p^*(\rho,z_i)=\p(0;\rho,z_i)$, which from Eq.(\ref{a2}) satisfy
 the following Klein-Gordon equation:
 \be
 \left(\Box_{\rho,z_i}-{m^*}^2\right)\p^*=0.
 \ee
 Since $-{m^*}^2=-\frac{D(D+2)}{4}<-\frac{(D-1)^2}{4}$, the scalar theory is not stable.
 The interacting theory
 in terms of $\pb=\p-\p_0$ is still well-defined and the corresponding
 conformally flat background is a $D$-dimensional de Sitter space which horizon is
 located at the singularity. It
 is interesting to note that ${m^*}^2=m^2\ell^2$, where $m$ is the
 mass of scalars $\pb$ on $dS_D$ with radius $\ell$, see appendix
 C.
 \section*{Acknowledgement}
 The author gratefully thanks  S. J. Rey and F. Shahbazi for useful
 discussions. The financial support of Isfahan University of
 Technology is acknowledged.
 \section*{Appendix A}
 In this appendix we show that $\p_0(x)$ is scale-invariant. We assume for
 simplicity that $D=4$. Under rescaling $x\to x'=\lambda x$ the scalar field changes as
 $\p(x)\to\p'(x')=\lambda^{-1}\p(x)$. By a scale-invariant object we mean a field that
 satisfies
 the relation $\delta_\epsilon\p=0$ where $\delta_\epsilon\p(x)=\p'(x)-\p(x)$ is the
 infinitesimal scale transformation  given by $\lambda=1+\epsilon$ for some
 infinitesimal $\epsilon$. To this aim we first note that under a general rescaling
 $\p(0)\to \p'(0)=\lambda^{-1}\p(0)$,
 thus, in~fact, $\p(x;\p(0))\to\p'(x;\p(0))=\lambda^{-1}\p(\lambda^{-1}x;\lambda\p(0))$.
 Defining $\beta^{-1}=\p(0)$, one can show that
 $\delta_\epsilon\p(x)=-\epsilon(1+x^I\partial_I)\p(x)$, where
 $x^I\in\{x^\mu,\beta\}.$ The $SO(D)$ invariant solutions of
 equation $\delta_\epsilon\p=0$ satisfying the condition
 $\p(0;\p(0))=\p(0)$ are
 $\p_k=\beta^{-1}\left(\frac{\beta}{\sqrt{\beta^2+x^2}}\right)^{k+2}$. It is easy
 to see that the action (\ref{action}) is invariant under the
 variation generated by $\delta_\epsilon$ and $\p_0$, among the others, is
 the solution of classical equation of motion.
 \section*{Appendix B}
 Here we give a detailed calculation of Hitchin
 information metric on the moduli space of $\p_0$
 (\ref{soliton}):
 \be
 \p_0=\left(\frac{D(D-2)}{g}\right)^{\frac{D-2}{4}}
 \left(\frac{\beta}{\beta^2+(x-a)^2}\right)^{\frac{D-2}{2}}.
 \ee
 From Eq.(\ref{L0}) one verifies that
 \be
 {\cal L}_0=\frac{g}{D}\left(\frac{D(D-2)}{g}\right)^{\frac{D}{2}}
 \left(\frac{\beta}{\beta^2+(x-a)^2}\right)^D.
 \ee
 Therefore
 \bea
 \partial_\beta\log{\cal
 L}_0&=&D\left(\frac{1}{\beta}-\frac{2\beta}{\beta^2+(x-a)^2}\right),\nn\\
 \partial_{a^i}\log{\cal L}_0&=&\frac{2D(x-a)_i}{\beta^2+(x-a)^2}.
 \eea
 Using these results and after some elementary calculations one
 can show that,
 \bea
 {\cal G}_{ij}&=&\frac{1}{N(D)}\int d^D x
 {\cal L}_0\partial_{a^i}\log{{\cal L}_0}\partial_{a^j}
 \log{{\cal L}_0}=\frac{4K(D)}{N(D)\beta^2}\delta_{ij}\int
 d^Dy\frac{y^2}{(1+y^2)^{D+2}},\nn\\
 {\cal G}_{\beta i}&=&{\cal G}_{i\beta}=\frac{1}{N(D)}
 \int d^D x {\cal L}_0\partial_{a^i}\log{{\cal L}_0}
 \partial_{\beta} \log{{\cal L}_0}=0,\nn\\
 {\cal G}_{\beta\beta}&=&\frac{1}{N(D)}
 \int d^D x {\cal L}_0\left(\partial_{\beta}\log{{\cal L}_0}
 \right)^2 =\frac{D K(D)}{N(D)\beta^2}\int
 d^Dy\frac{1}{(1+y^2)^D}\left(1-\frac{2}{(1+y^2)}\right)^2.
 \eea
 where $K(D)=g^{1-D/2}D^{D/2+1}(D-2)^{D/2}$.
 By performing the integrations and using Eq.(\ref{N}), one obtains,
 \be
 {\cal G}_{IJ}=\frac{1}{\beta^2}\delta_{IJ},
 \ee
 in which $\delta_{IJ}=1$ if $I=J$ and vanishes otherwise.
 \section*{Appendix C}
 In this appendix we briefly review free scalar field theory in $D+1$
 dimensional (Euclidean) curved space-time \cite{Ted} and say few words about the
 classical geometry of de Sitter space \cite{Les}. The action for the scalar field $\phi$ is
 \be
 S=\int d^Dx\;
 \sqrt{\left|g\right|}\frac{1}{2}\left(g^{\mu\nu}\partial_\mu\phi\partial_\nu\phi+(m^2+\xi
 R)\phi^2\right),
 \ee
 for which the equation of motion is
 \be
 \left(\Box - m^2-\xi R\right)\phi=0,\hspace{1cm}
 \Box\p=|g|^{-1/2}\partial_\mu\left(\left|g\right|^{1/2}g^{\mu\nu}\partial_\nu\p\right).
 \ee
 (With $\hbar$ explicit, the mass $m$ should be replaced by
 $m/\hbar$.) The case with $m=0$ and $\xi=\frac{D-2}{4(D-1)}$ is referred to
 as conformal coupling.
 \par
 The curvature tensor $R^\mu_{\ \nu\rho\sigma}$ in term of Levi-Civita
 connection,
 \be
 \Gamma^\mu_{\
 \nu\rho}=\frac{1}{2}g^{\mu\alpha}(\partial_\rho g_{\alpha\nu}+\partial_\nu
 g_{\alpha\rho}-\partial_\alpha g_{\nu\rho}),
 \ee
 is given as follows,
 \be
 R^\mu_{\ \nu\rho\sigma}=\partial_\rho\Gamma^\mu_{\ \nu\sigma}-
 \partial_\nu\Gamma^\mu_{\ \rho\sigma}+\Gamma^\mu_{\ \rho\alpha}\Gamma^\alpha_{\
 \nu\sigma}-\Gamma^\mu_{\ \alpha\sigma}\Gamma^\alpha_{\ \nu\rho}.
 \ee
 The Ricci tensor $R_{\nu\sigma}=R^\mu_{\ \nu\mu\sigma}$ and the
 curvature scalar $R=g^{\nu\sigma}R_{\nu\sigma}$.
 \par
 The metric of a conformally flat space-time can be given as
 $g_{\mu\nu}=\Omega\delta_{\mu\nu}$, where $\Omega$ is some
 function of space-time coordinates. One can easily show that,
 \bea
 R_{\mu\nu}&=&\frac{2-D}{2}\partial_\mu\partial_\nu(\log{\Omega})-\frac{1}{2}\delta_{\mu\nu}
 \nabla^2(\log{\Omega})\nn\\&+&\frac{D-2}{4}
 \left(\partial_\mu(\log{\Omega})\partial_\nu(\log{\Omega})-\delta_{\mu\nu}
 \delta^{\rho\sigma}\partial_\rho(\log{\Omega})\partial_\sigma(\log{\Omega})\right)\nn\\
 \Omega R&=&(1-D)\nabla^2(\log{\Omega})+\frac{(1-D)(D-2)}{4}
 \delta^{\mu\nu}\partial_\mu(\log{\Omega})\partial_\nu(\log{\Omega}).
 \eea
 By inserting $\pp=\Omega^{\frac{D-2}{4}}\pb$ in the action
 $S[\pp]=\int d^Dx \frac{1}{2}\delta^{\mu\nu}\partial_\mu\pp\partial_\nu\pp$,
 one obtains,
 \bea
 S[\pp]&=&\int d^Dx \left(\frac{1}{2}\Omega^{\frac{D-2}{2}}
 \delta^{\mu\nu}\partial_\mu\pb\partial_\nu\pb-
 \frac{1}{2}\left(\Omega^{\frac{D-2}{4}}\nabla^2
 \Omega^{\frac{D-2}{4}}\right)\pb^2\right)\nn\\
 &=&\int d^Dx \sqrt{g}\left(\frac{1}{2}g^{\mu\nu}\partial_\mu\pb\partial_\nu\pb+
 \frac{1}{2}\xi R\pb^2
 \right).
 \eea
 To obtain the last equality the identities
 $g_{\mu\nu}=\Omega\delta_{\mu\nu}$ and $\xi\sqrt{g}R=
 -\Omega^{\frac{D-2}{4}}\nabla^2
 \Omega^{\frac{D-2}{4}}$ are used. Consequently the free massless scalar
 theory on $D$-dimensional Euclidean space, is (classically)
 equivalent to some conformally coupled scalar theory on the
 corresponding conformally flat background.
 \par
 A $D$-dimensional de Sitter (dS) space may be realized as the
 hypersurface described by the equation
 $-X_0^2+X_1^2+\cdots+X_D^2=\ell^2$. $\ell$ is called the de
 Sitter radius. By replacing $\ell^2$ with $-\ell^2$ the
 hypersurface is the $D$-dimensional anti de Sitter (AdS) space.
 (A)dS spaces are Einstein manifolds with positive (negative)
 scalar curvature. The Einstein metric
 $G_{\mu\nu}=R_{\mu\nu}-\frac{1}{2}Rg_{\mu\nu}$, satisfies
 $G_{\mu\nu}+\Lambda g_{\mu\nu}=0$, where $\Lambda=\frac{(D-2)(D-1)}{2\ell^2}$ is the
 cosmological constant. From Eq.(\ref{Lambda}) one obtains
 $-4m^2\ell^2=D(D+2)$ which determines the radius of the $\mbox{dS}_D$ background
 in terms of the mass of scalar field $\pb$.
  

\newpage
\begin{thebibliography}{99}
 \bibitem{Mal} J. M. Maldacena, Adv. Theor. Math. Phys. 2, (1998) 231, hep-th/9711200.
 \bibitem{Witten} E. Witten, Adv.Theor.Math.Phys. 2 (1998) 253, hep-th/9802150.
  \bibitem{Reh} M. Duetsch and K. H. Rehren, Lett.Math.Phys. 62 (2002)
 171-184, hep-th/0204123.
 \bibitem{Strom} A. Strominger, JHEP 0110 (2001) 034, hep-th/0106113.
 \bibitem{map} F. Loran, Phys.Lett. B601 (2004) 192-196,
 hep-th/0404067.
 \bibitem{Hen}  M. Henneaux, {\em Boundary terms in the AdS/CFT correspondence
 for spinor fields}, hep-th/9902137;\\
 W. M\"{u}ck and K. S. Viswanathan, Phys.Rev. D58 (1998) 106006,
 hep-th/9805145;\\
 M. Henningson and K. Sfetsos, Phys.Lett.  B431 (1998)
 63-68, hep-th/9803251;\\
 G. E. Arutyunov and S. A. Frolov, Nucl.Phys. B544 (1999)
 576-589,hep-th/9806216;\\
  A.M. Ghezelbash, K. Kaviani, S. Parvizi and A.H.
 Fatollahi, Phys.Lett. B435 (1998) 291-298, hep-th/9805162.
 \bibitem{Spin} F. Loran, JHEP06(2004)054, hep-th/0404135.
 \bibitem{phi4} F. Loran, Phys.Lett. B605 (2005) 169-180,
 hep-th/0409267.
 \bibitem{Hit} N. J. Hitchin, {\em The Geometry and Topolgy on
 Moduli Spaces} in {\em Global Geometry and Mathematical Physics}
 1451, (Springer, Heidelberg, 1988) 1-48.
 \bibitem{Blau} M. Blau, K. S. Narain and G. Thompson, {\em Instantons, the
 Information Metric, and the AdS/CFT Correspondence},
 hep-th/0108122.
 \bibitem{Rey}  R. Britto, B. Feng, O. Lunin and S. J. Rey, Phys.Rev. D69 (2004)
 126004, hep-th/0311275;\\
 S. Parvizi, Mod.Phys.Lett. A17 (2002) 341-354, hep-th/0202025.
 \bibitem{Sol}J. de Boer and S. N. Solodukhin, Nucl.Phys. B665 (2003)
 545-593, hep-th/0303006;\\
 S. N. Solodukhin, "{\em Reconstructing Minkowski Space-Time}, hep-th/0405252.
 \bibitem{Ted}  T. Jacobson, {\em Introduction to Quantum Fields in Curved Spacetime and
the Hawking Effect}, gr-qc/0308048.
\bibitem{Les} M. Spradlin, A. Strominger and A. Volovich,
{\em Les Houches Lectures on De Sitter Space}, hep-th/0110007.
\end{thebibliography}
\end{document}